\documentclass[aps, prb, preprint, superscriptaddress]{revtex4}
\usepackage{graphicx}
\usepackage{dcolumn}
\usepackage{bm}
\usepackage{natbib}
\usepackage{upgreek}
\usepackage{amsmath}
\usepackage{amssymb}

\newcommand*{\Sm}{Sm$_{0.5}$La$_{0.5}$Fe$_3$(BO$_3$)$_4$}
\newcommand*{\vect}[1]{\mathbf{#1}}

\begin{document}

\title{Large directional anisotropy in multiferroic ferroborate}

\author{A. M. Kuzmenko}
\affiliation{Prokhorov General Physics Institute, Russian Academy of
Sciences, 119991 Moscow, Russia}
\author{V. Dziom}
\author{A. Shuvaev}
\author{Anna Pimenov}
\author{M. Schiebl}
\affiliation{Institute of Solid State Physics, Vienna University of
Technology, 1040 Vienna, Austria}
\author{A. A. Mukhin}
\author{V. Yu. Ivanov}
\affiliation{Prokhorov General Physics Institute, Russian Academy of
Sciences, 119991 Moscow, Russia}
\author{I. A. Gudim}
\author{L. N. Bezmaternykh}
\affiliation{L.V. Kirensky Institute of Physics Siberian Branch of RAS,
660036 Krasnoyarsk, Russia}
\author{A. Pimenov}
\affiliation{Institute of Solid State Physics, Vienna University of
Technology, 1040 Vienna, Austria}

\begin{abstract}

One of the most fascinating and counter-intuitive recent effects in multiferroics is the directional anisotropy, the asymmetry of light propagation with respect to the direction of propagation. In such case the absorption in a material can be different for opposite directions. Beside absorption, different velocities of light for different directions of propagation may be also expected, which is termed directional birefringence.
In this work, we demonstrate large directional anisotropy in a multiferroic samarium ferroborate. The effect is observed for linear polarization of light in the range of millimeter-wavelengths, and it survives down to low frequencies. The dispersion and absorption close to the electromagnon resonance can be controlled by external magnetic field  and is fully suppressed in one direction. By changing the geometry of the external field, samarium ferroborate shows giant optical activity, which makes this material to a universal tool for optical control: with a magnetic field as an external parameter it allows switching between two functionalities: polarization rotation and directional anisotropy.

\end{abstract}

\date{\today}


\maketitle

\section{Introduction}

Multiferroics are materials which exhibit electric and magnetic order simultaneously~\cite{fiebig_jpd_2005, eerenstein_nature_2006, tokura_jmmm_2007, wang_advph_2009}. Due to the coupling of electric and magnetic effects, these materials show a strong potential to control electricity and magnetism and, more generally, the properties and propagation of light. An unusual way to influence the light beam in a material is provided by directional anisotropy. This effect is the inequivalence of forward and backward directions of propagation and it may be divided into directional dichroism and directional birefringence. Directional dichroism results in an asymmetry of the absorption coefficient. Directional birefringence may be defined as different velocity of light for forward/backward direction, or, equivalently, by difference in the refractive index.

One way to obtain the asymmetric transmission is a material is given by the magnetochiral anisotropy \cite{barron_mph_1984}. This effect is physically close to the Faraday rotation \cite{zvezdin_book} and results in non-equivalent transmission for light propagating parallel or antiparallel to the magnetisation of the sample. Experimental demonstrations of magnetochiral anisotropy exist in a series of systems \cite{rikken_pre_1998, krstic_jcp_2002, sautenkov_prl_2005} including ferromagnets\cite{train_nm_2008} and multiferroics\cite{bordacs_nphys_2012, kibayashi_nc_2014,kezsmarki_nc_2014, okamura_prl_2015}.

Further possibility to break the symmetry with respect to propagation direction may be realized in multiferroics and with magnetization perpendicular to the light propagation. In multiferroics with electric polarization $\mathbf{P}$ perpendicular to the magnetization $\mathbf{M}$ a nonzero toroidal moment~\cite{fiebig_jpd_2005, tokura_jmmm_2007} appears, $\mathbf{T}=\mathbf{P}\times \mathbf{M}$,  which may be parallel or antiparallel to the propagation direction. Directional anisotropy based on toroidal moment has been recently demonstrated in multiferroics~\cite{kezsmarki_prl_2011, takahashi_nphys_2012, takahashi_prl_2013} and is realised in present work.

More strictly, directional anisotropy may be defined in terms of  non-reciprocity and is equivalent to an asymmetry with respect to interchanging the source and detector. Mathematically, non-reciprocal effects in optics can by quantified via a property called reaction~\cite{kong_book} which measures the difference between backward and forward geometries. In several specific cases, the reciprocity (i.e. equivalence of forward and backward directions) can be proven rigorously, e.g. for scalar fields~\cite{born_book} or for media with symmetric susceptibilities and without magnetoelectric terms~\cite{landau_book8}. A recent review about the topic may be found in Ref.~[\onlinecite{potton_rpp_2004}]. Earlier results on non-reciprocal optics~\cite{krichevtsov_jetp_1988, krichevtsov_jpcm_1993, mukhin_ferr_1997} have been obtained in a classical magnetoelectric material Cr$_2$O$_3$.

The propagation of light in the matter is governed by the material equations:
\begin{equation}
\begin{array}{c}
 \vect{B} = \mu_0\hat{\mu} \vect{H} + \sqrt{\varepsilon_0 \mu_0}\hat{\chi}^{me} \vect{E}, \\
 \vect{D} = \sqrt{\varepsilon_0 \mu_0} \hat{\chi}^{em} \vect{H} + \varepsilon_0 \hat{\varepsilon} \vect{E}. \\
\end{array}
\label{constitutive_equations}
\end{equation}
Here $\vect{B}$ is the magnetic flux density, $\vect{D}$ is the electric displacement field; $\vect{E}$ and $\vect{H}$ are the electric and magnetic fields; $\hat{\varepsilon}$, $\hat{\mu}$ and $\hat{\chi}_{me/em}$ are the matrices of electric permittivity, magnetic permeability and magnetoelectric susceptibilities, respectively.
As may be shown by rigorous calculation \cite{kong_book, tretyakov_book_2001} the necessary conditions for reciprocal propagation are given by
\begin{equation}\label{eqreciprocal}
  \hat{\varepsilon} = \hat{\varepsilon}^T, \quad \hat{\mu} = \hat{\mu}^T, \quad \hat{\chi}_{me} = -\hat{\chi}_{em}^T \quad .
\end{equation}
Here $()^T$ denotes the transposed matrix.

One well-known case of a non-reciprocal transmission is given by the Faraday rotation. Here the effect is realised by antisymmetric off-diagonal elements of permittivity or permeability. As will be shown later, in case of samarium ferroborate the non-reciprocity is due to symmetric off-diagonal terms in the magnetoelectric susceptibility ($\chi_{xy}^{me}$).

An instructive question within the scope of the present work is: What is the symmetry operation exchanging the source and detector, i.e. which operation reverses the direction of light? If we consider plane electromagnetic waves as a standard experimental tool, both time inversion ($t\rightarrow -t$) and space inversion ($r \rightarrow -r$) would reverse the propagation direction. In case when linearly polarized waves are the eigenwaves of the problem (as in experiments below), both time and space inversions exchange the direction but the linear polarization of the wave is preserved. The situation gets significantly different if circular (or elliptically) polarized waves are the propagating eigenmodes in the sample. We recall that besides changing the propagation direction, space and time inversions differently influence the fields of the electromagnetic wave: time inversion preserves the direction of the electric field, but inverts the magnetic field. The space inversion does just the opposite: the electric field is inverted and the magnetic field is preserved. This difference has fundamental consequence for the circular polarization. As may be easily shown, the space inversion interchange clockwise and counterclockwise polarizations (i.e. the handedness), but the time inversion does not. From the definition of reciprocity it can be shown that the rotation sense of circular waves must be kept after inverting the propagation direction. Therefore, in practical experiments with inverted propagation of light, space inversion symmetry does not seem to be adequate at least in case of circular (or elliptical) polarizations.

The propagation of light within a medium is characterized by a wave vector and - for a given propagation direction - may be reduced to a complex refractive index. The real part of the refractive index is responsible for the collection of a phase shift during the propagation. In other words, it reflects the ratio between the light velocity in the media and in vacuum. The imaginary part of the refractive index is responsible for the energy loss and it is termed absorption coefficient. Directional anisotropy of the absorption coefficient is called directional dichroism. Strong directional dichroism in multiferroics arises due to coupling of electric and magnetic order~\cite{kezsmarki_prl_2011, takahashi_nphys_2012,takahashi_prl_2013, kezsmarki_nc_2014, kibayashi_nc_2014}. In addition, specific symmetry conditions~\cite{szaller_prb_2013} should be fulfilled in such experiments, e.g. the orthogonal mutual orientation of magnetization, electric polarization, and propagation direction.

Electromagnetic waves are solutions of Maxwell equations, where six components of electric and magnetic field and the wave vector (propagation vector) are unknowns. Using two Maxwell equations, $\nabla \mathbf{D}= 0$ and  $\nabla \mathbf{B}= 0$, two field components can be removed from the system. Here $\mathbf{D}$ and $\mathbf{B}$ are electric displacement and magnetic flux density, respectively. Using boundary conditions, the problem reduces to a four-dimensional linear problem for eigenvalues and eigenvectors. Four possible eigenvalues represent four possible propagation constants in the media. In general case, all eigenvalues are different. However, due to inversion symmetry in the majority of the problems, two pairs of solutions are equal in absolute values and opposite in sign. They correspond to two possible polarizations of the electromagnetic wave in the sample with two (forward and backward) propagation directions for each wave. Non-reciprocal effects arise when the eigenvectors are different for opposite directions.

Here we present a material that can be considered as a universal tool for the control of the optical properties. Using solely an external magnetic field as a parameter two different effects may be switched and modified: (i) giant optical activity \cite{kuzmenko_prb_2014} (polarization rotation) and (ii) giant directional birefringence and dichroism for linearly polarized light. For intermediate orientations both effects are mixed realizing a general case of four-colored transmission~\cite{kezsmarki_nc_2014}.

Magnetoelectric rare-earth ferroborates~\cite{vasiliev_ltp_2006,kadomtseva_ltp_2010} RFe$_3$(BO$_3$)$_4$ (R = rare earth ion) represent a newly discovered material class with strong magnetoelectric coupling.
Especially in ferroborates with R = Sm, Ho colossal magnetic
field-induced changes in the dielectric constant have been observed
\cite{mukhin_jetpl_2011,chaudhury_prb_2009}. In the case of \Sm~ it has been proven experimentally
that an intrinsic magnetoelectric excitation are
responsible for the observed effects
\cite{kuzmenko_prb_2014, mukhin_jetpl_2011,kuzmenko_jetp_2011}. Such excitations in
magnetoelectric materials are called electromagnons
\cite{pimenov_jpcm_2008,sushkov_jpcm_2008}, and they are defined as
magnetic excitations that interact with the electric component of
electromagnetic radiation.

\section{Experimental details}

Spectroscopic experiments in the terahertz frequency range (40~GHz
$< \nu <$ 1000 GHz) have been carried out in a Mach-Zehnder
interferometer arrangement~\cite{volkov_infrared_1985} which allows
measurements of the amplitude and the phase shift in a geometry with
controlled polarization of radiation. Theoretical transmittance
curves \cite{shuvaev_sst_2012} for various geometries were
calculated from the susceptibilities within the Berreman
formalism \cite{berreman_josa_1972}. The experiments in external
magnetic fields up to 7~T have been performed in a superconducting
split-coil magnet with polypropylene windows. Static dielectric
measurements have been done using commercial impedance analyzer
equipped with a superconducting magnet.

Large single crystals of \Sm, with typical dimensions of $\sim 1$
cm, have been grown by crystallization from the melt on seed
crystals.

\begin{figure}[tbp]
\begin{center}
\includegraphics[width=0.85\linewidth, clip]{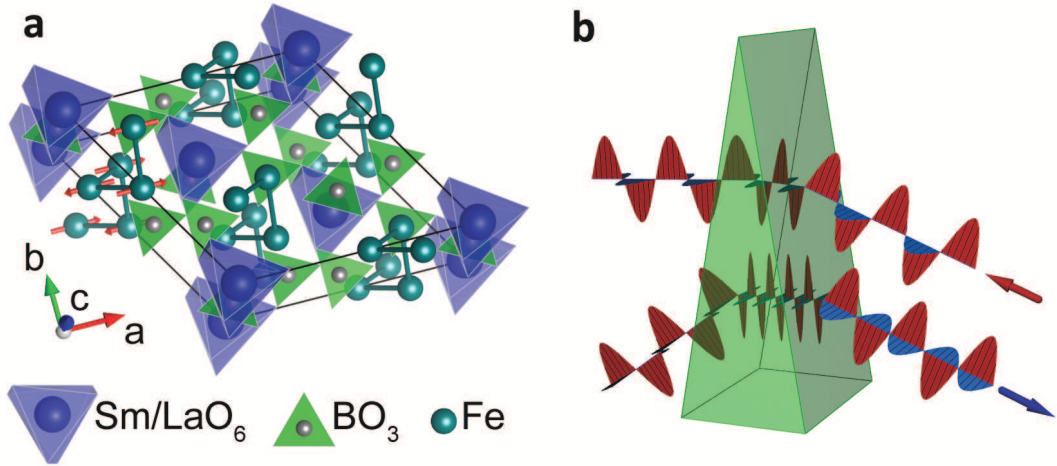}
\end{center}
\caption{\emph{Crystal structure of \Sm ~ and directional birefringence.} \textbf{a} - Basic
structural elements in \Sm. The direction of Fe-spins is given in the geometry with external magnetic field parallel to the $b$-axis. \textbf{b} - Definition of directional birefringence. The backward beam (red) is only weakly refracted by the sample.} \label{fig1}
\end{figure}

\section{Results}
\subsection{Samarium Ferroborate}
The material used in the present study is \Sm. This compound is closely similar to the pure samarium ferroborate~\cite{vasiliev_ltp_2006, kadomtseva_jetp_2012}, SmFe$_3$(BO$_3$)$_4$. Doping with lanthanum has an advantage of suppressing the growth of domains with different symmetry. The presence of domains in the sample would reduce the value of the magnetoelectric non-reciprocal effects. The crystallographic structure of \Sm~ is shown in
Fig. 1\textbf{a}. It contains two interacting subsystems given by
Sm$^{3+}$/La$^{3+}$ and Fe$^{3+}$ ions. The iron subsystem orders
antiferromagnetically below $T_N=34~ K$ with an easy-plane magnetic
structure oriented perpendicularly to the trigonal c-axis. Although
the Sm$^{3+}$ moments play an important role in the magnetoelectric
properties of \Sm, they probably do not order up to the lowest
temperatures.

Static electric polarization in multiferroic ferroborates can be
explained by symmetry arguments and by taking into account that
Fe$^{3+}$ moments are oriented antiferromagnetically within the
crystallographic
$ab$-plane~\cite{zvezdin_jetpl_2005,zvezdin_jetpl_2006}. Within the
topic of the present work the terms governing the ferroelectric
polarization along the $a$ and $b$-axis (or $x$ and $y$-axes) are of basic importance. For the R32 space group of
borates these terms are given by
\begin{equation}\label{p}
P_x \sim L_x^2-L_y^2 ,\quad  P_y \sim -2L_xL_y \ .
\end{equation}
Here $\vect{L}=\mathbf{M}_1-\mathbf{M}_2$ is the antiferromagnetic vector with $\mathbf{M_1}$ and
$\mathbf{M_2}$ being the magnetic moments of two (antiferro-)magnetic
Fe$^{3+}$ sublattices. Further details of the symmetry analysis of the
static magnetoelectric effects in \Sm~ can be found in Appendix and in Refs.~
[\onlinecite{zvezdin_jetpl_2005,zvezdin_jetpl_2006,popov_prb_2013}].

The simple expression Eq.~(\ref{p}) allows to understand the behavior of
static and dynamic properties in external magnetic fields.
For magnetic fields along the
$a(x)$-axis an antiparallel orientation of electric polarization with
respect to the $a$-axis is stabilized. In this case \Sm~ reveal strong electromagnon mode~\cite{kuzmenko_prb_2014, pimenov_jpcm_2008} which can be excited separately via electric, ${e}\|b(y)$, or magnetic, ${h}\|a(x)$, channels. Here ${h}$ and ${e}$ are the ac-components of the electromagnetic radiation. As has been shown recently~\cite{kuzmenko_prb_2014}, in the geometry $\mathbf{B}\|a$ samarium ferroborate exhibits strong optical activity, i.e. the polarization of the incident radiation can be rotated on a controllable way.

\subsection{Non-reciprocal transmission}

The geometry with magnetic field along the $b(y)$-axis is very promising from the point of view of non-reciprocal effects. External field $\mathbf{B}\|b$
stabilizes the magnetic configuration with the magnetic moments oriented along the $a$-axis ($L_y=0$, $L_x\neq 0$).
In agreement with Eq.~(\ref{p}), the static polarization
is oriented parallel to the $a$-axis. The electromagnon is excited simultaneously via the electric and magnetic channel: ${h}\|a(x)$, ${e}\|b(y)$. In this case the magnetoelectric coupling starts to distinguish between two possible propagation directions. This effect arises because electric polarization, $\mathbf{P}\|a$, is oriented perpendicular to the induced magnetization $\mathbf{M}\|\mathbf{B}\|b$. Consequently, \Sm~ gets a toroidal~\cite{fiebig_jpd_2005} moment, $\mathbf{T}=\mathbf{P}\times \mathbf{M}$, which may be parallel or antiparallel to the propagation direction~\cite{tokura_jmmm_2007}. The toroidal moment allows the existence of strong directional birefringence in \Sm.

The susceptibility matrices in Eq.~(\ref{constitutive_equations}) $\hat{\chi}^m(\omega), \hat{\chi}^{me}(\omega), \hat{\chi}^{em}(\omega), \hat{\chi}^e(\omega)$ can be written in a simplified form (see Appendix):
\begin{equation}
\begin{array}{cc}
\hat{\chi}^m(\omega) =
\left( \begin{array}{ccc}
\chi_{xx}^m & 0 & \chi_{xz}^m \\
0 & \chi_{yy}^m & 0 \\
-\chi_{xz}^m & 0 & \chi_{xx}^m \\
\end{array} \right) & \hat{\chi}^{me}(\omega) =
\left( \begin{array}{ccc}
0 & \chi_{xy}^{me} & 0 \\
0 & 0 & 0 \\
0 & \chi_{zy}^{me} & 0 \\
\end{array} \right) \\[3em]
\hat{\chi}^{em}(\omega) = \left( \begin{array}{ccc}
0 & 0 & 0 \\
\chi_{xy}^{me} & 0 & -\chi_{zy}^{me} \\
0 & 0 & 0 \\
\end{array} \right) & \hat{\chi}^e(\omega) = \left( \begin{array}{ccc}
\chi_{xx}^e & 0 & 0 \\
0 & \chi_{yy}^e & 0 \\
0 & 0 & \chi_{zz}^e \\
\end{array} \right). \\
\end{array}
\label{susceptibilities}
\end{equation}

As shown in more details in the Appendix, for an electromagnetic wave propagating along the $z$-direction the complex refractive indexes of the forward ($n^+$) and backward ($n^-$) solutions are different:
\begin{equation}\label{eqnk}
  n^{\pm}= \sqrt{\tilde{\varepsilon}_y \tilde{\mu}_x} \pm \tilde{\alpha}_{xy} \ .
\end{equation}

Here $\tilde{\alpha}_{xy} = \chi_{xy}^{me} -\chi_{zy}^{me} \chi_{xz}^{m} / \mu_{z}$, $\tilde{\varepsilon}_y = \varepsilon_y + (\chi_{zy}^{me})^2/\mu_{z}$, and $\tilde{\mu}_x = \mu_x + (\chi_{xz}^{m})^2/\mu_{z}$ are the renormalized material parameters;
$\varepsilon_y = 1+ \chi_{yy}^e$ is the dielectric permittivity along the $y$-axis, $\mu_x= 1+ \chi_{xx}^m$ and $\mu_z= 1+ \chi_{zz}^m$ are the magnetic permeabilities along the $x$-axis and $z$-axis, respectively. In present approximation $\mu_x \approx \mu_z$. The solutions in Eq.~(\ref{eqnk}) are written for a linearly polarized electromagnetic wave with a polarization $e \| y$, $h\| x$. The solution for the perpendicular polarization with $e \| x$, $h\| y$ is trivial and can be written as $n^{\pm}= \sqrt{\varepsilon_x \mu_y}$. Importantly, two possible solutions for the propagating wave are linearly polarized. Therefore, no polarization rotation is expected for the geometry with $\mathbf{B}\|b$ and the two linear polarizations do not mix. This is in contrast to a related geometry with $\mathbf{B}\|a$ revealing optical activity~\cite{kuzmenko_prb_2014}.

\begin{figure}[tbp]
\begin{center}
\includegraphics[angle=270, width=0.85\linewidth, clip]{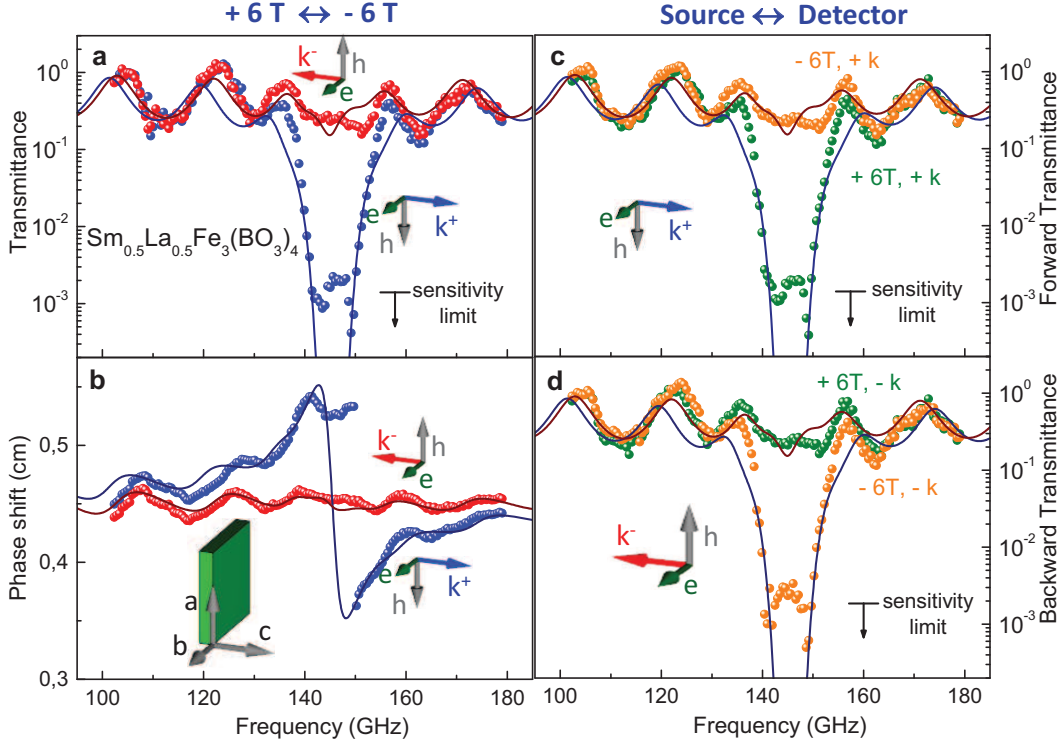}
\end{center}
\caption{\emph{Non-reciprocal transmission in \Sm.}
Left panels: non-reciprocity is achieved by the inversion of magnetic field. \textbf{a} - Forward/backward transmission. \textbf{b} - Non-reciprocal phase shift, $\varphi / \nu$. Right panels: non-reciprocity is achieved by explicit exchange of source and detector. \textbf{c} - Forward transmission for two opposite directions of external magnetic field. \textbf{d} - Backward transmission. The geometry of the experiment is given in the inset. Symbols - experiment, solid lines - fits according to the Berreman model as described in the Appendix.} \label{figtr}
\end{figure}

Typical transmittance spectra in the geometry with non-reciprocal effects ($\mathbf{B}\|b$) in the frequency range of our
spectrometer are shown in Fig.~\ref{figtr}. Switching between forward and backward geometry is experimentally achieved by inverting the direction of the external magnetic field, i.e. the direction of the induced magnetization, $\mathbf{M}\|\mathbf{B}$. We recall that in agreement with Eq.~(\ref{p}) the direction of electric polarization does not change, but the toroidal moment $\mathbf{M}\times \mathbf{P}$ does. Therefore, the inversion of the external magnetic field is equivalent to the inversion of the propagation direction. From the point of view of symmetry, two experimentally relevant cases are: $\mathbf{M}\times \mathbf{P} \uparrow \uparrow \mathbf{k}$ and $\mathbf{M}\times \mathbf{P} \uparrow \downarrow \mathbf{k}$.

In order to check explicitly, that the exchange of the source and detector does lead to the same results, the corresponding experiments have been carried out. The results of these experiments are presented in the right panels of Fig.~\ref{figtr}. Only the absolute values of the transmission have been measured in this case as the Mach-Zehnder interferometer~\cite{volkov_infrared_1985} can not be easily inverted. Fig.~\ref{figtr}\textbf{c} shows the transmittance in the geometry with forward propagation direction and with two opposite values of external magnetic field. We state that similarly to the spectra in the left panels, the electromagnon is strong for one direction of the magnetic field only. The transmittance spectra for the backward direction are presented in Fig.~\ref{figtr}\textbf{d}. Here the situation is just the opposite to that in the panel \textbf{c}: the field values with excited and silent electromagnon interchange.

We note that no Faraday rotation is expected in present case, because the Voigt geometry with $\mathbf{B}\|b$, $\vect{k}\|c$ has been used.

Figs.~\ref{figtr}\textbf{a,c,d} demonstrate the dependence of the transmission intensity on the propagation direction, which is equivalent to a directional dichroism. In \Sm~ the directional dichroism is observed only close to the electromagnon resonance frequency, $\nu_0 \approx 146$~GHz ($B = 6$~T). Already for frequencies $\pm 10$~GHz apart from the resonance, forward and backward transmission intensities coincide within experimental accuracy.

\subsection{Directional birefringence}

Figure~2\textbf{b} shows the difference in phase shift (optical thickness) in \Sm~ for forward/backward propagation. Again, close to the electromagnon frequency strong difference in optical thickness is observed between two directions. An important contrast to the asymmetric transmission in Fig.~2\textbf{a} is a broadband character of the observed effect, i.e. substantial variation in optical thickness is detected in the full frequency range of the present experiment.

\begin{figure}[tbp]
\begin{center}
\includegraphics[width=0.6\linewidth, clip]{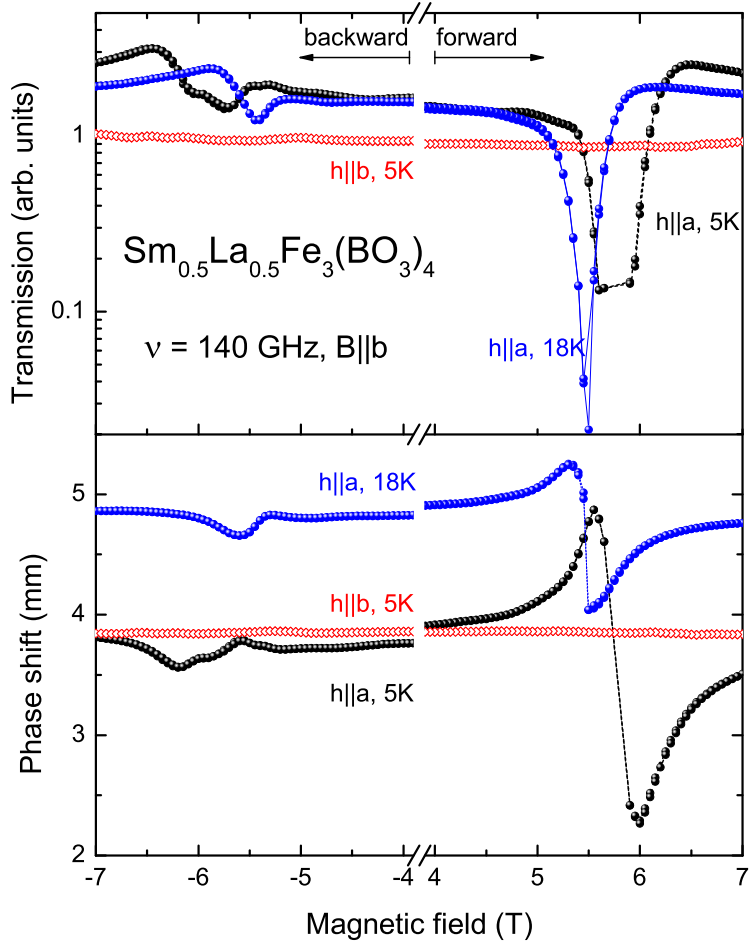}
\end{center}
\caption{\emph{Absence of polarization rotation}.
Magnetic field dependence of the transmittance spectra in \Sm. Positive/negative values of magnetic field is equivalent to forward/backward direction of propagation, respectively. \textbf{a} - Amplitude of the transmitted signal. \textbf{b} - Relative changes of the optical thickness. Open symbols (blue and black) - magnetoelectric geometry of the experiment with both, electric and magnetic channels, excited. Full symbols - silent geometry: the electromagnon is not excited. Solid lines - fits as described in the Appendix. Absence of the electromagnon excitation in the silent geometry ($\mathbf{B}\|h\|b$) demonstrate zero polarization rotation.} \label{figfld}
\end{figure}
Figure \ref{figfld} demonstrates that two solutions for the propagating waves are linear and do not mix. In the geometry with $h\|a$ clear electromagnon excitation is observed close to $\pm 6$ Tesla, as in this configuration electric and magnetic excitation channels are active. We note here again the contrast between positive and negative directions of the magnetic field. In a simple interpretation using Eq.~(\ref{eqnk}) for the propagation constants, one can say that for positive fields the magnetoelectric contribution is added to the "conventional" electromagnon mode. For negative magnetic fields the magnetoelectric susceptibility is subtracted from the refractive index. The characteristic values of the susceptibilities are close to typical numbers, for which the electromagnon is nearly suppressed in the backward direction.

The silent geometry in Fig.~\ref{figfld} with $\mathbf{B}\|{h}\|b$, ${e}\|a$ shows no excitation in the spectra. This observation supports the simple model presented above, in which no electromagnon and no magnetoelectric contribution is present. Most importantly, this demonstrates the absence of the polarization rotation in the configuration $\mathbf{B}\|b$, $\vect{k}\|c$. That is, active and silent polarizations do not mix and remain linear. The observed effect remains correct as long as the spectra are dominated by a low-frequency electromagnon. The effect of the high-frequency electromagnon \cite{kuzmenko_jetp_2011,kuzmenko_jetpl_2011} at about 320~GHz can be neglected. As may be expected, if the interaction of both electromagnons is substantial, certain effects of the polarization rotation would appear. 
\begin{figure}[tbp]
\begin{center}
\includegraphics[width=0.85\linewidth, clip]{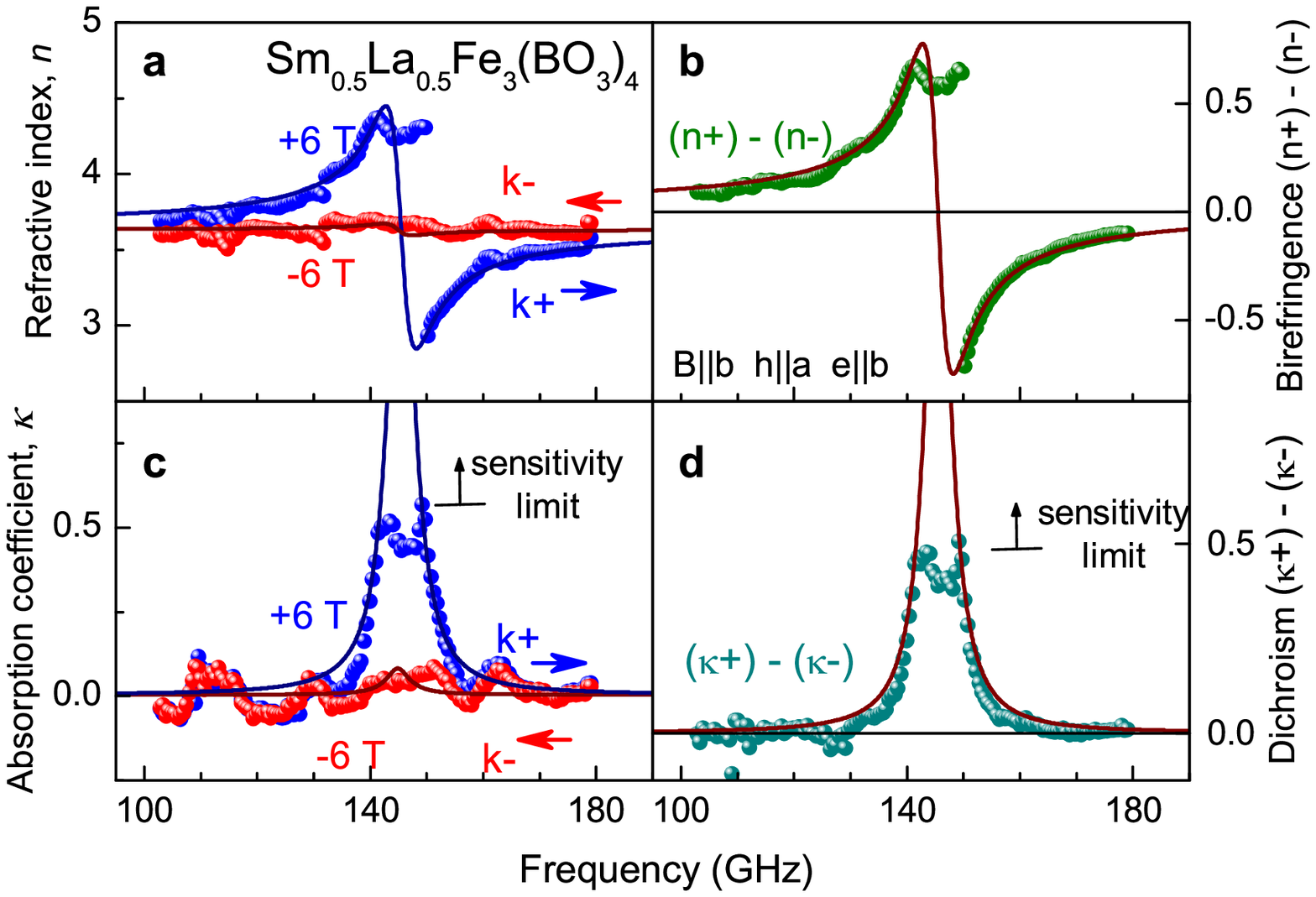}
\end{center}
\caption{\emph{Directional birefringence and dichroism in \Sm}.
\textbf{a} - Refractive index in \Sm~ for forward (blue) and backward (red) propagation of the millimeter-wave radiation. \textbf{b} - Directional birefringence. \textbf{c} - Absorption coefficient in \Sm~ for forward and backward propagation. \textbf{d} - Directional dichroism. Symbols -
experiment, solid line - fits according to the model described in the text. } \label{fignk}
\end{figure}

\section{Discussion}

Figure \ref{fignk} summarizes the central results of this work, i.e. directional birefringence and directional dichroism in \Sm. As described in more details in the Appendix, the refractive index ($n$) and the absorption coefficient ($\kappa$) have been obtained from the transmission and phase shift spectra shown in Fig.~\ref{figtr}. As the formulas do include the interferences within the sample surfaces, Fabry-P\'{e}rot-like oscillations seen in the transmittance spectra (Fig.~\ref{figtr}) are substantially suppressed in the data of Fig.~\ref{fignk}. As expected from the measured spectra of the optical thickness, the refractive index in \Sm~ is strongly direction-dependent. Moreover, a substantial difference in $n$ is present at all frequencies shown in Fig.~\ref{fignk}. The difference is also clearly seen in Fig.~\ref{fignk}\textbf{b} which directly presents the directional birefringence (i.e. $\Delta n = n^+ - n^-$). Because the electromagnon frequency strongly depends upon an external magnetic field, $\nu_0 \sim B$, the birefringence can be varied by magnetic field and even changes sign on crossing the resonance frequency.

In agreement with the data in Fig.~\ref{fignk}\textbf{b}, the directional dichroism remains nonzero even at static frequencies. According to Eq.~(\ref{eqnk}) the static value is
\begin{equation}\label{eqst}
  \Delta n (0) = n^+(0) - n^-(0) \approx 2\chi_{xy}^{me}(0) \ .
\end{equation}
We note that the static value of the refractive index is ill-defined with increasing wavelength and Eq.~(\ref{eqst}) is probably relevant down to microwaves only.

In a contrast to the directional birefringence, the directional dichroism in Sm$_{0.5}$La$_{0.5}$Fe$_3$(BO$_3$)$_4$ is limited in frequency (see Fig.~\ref{fignk}\textbf{c,d}). Although the dichroism is strong close to the electromagnon frequency, is gets unmeasurably small $\pm 10$~GHz apart from the resonance frequency.

A remarkable result seen in Fig.~\ref{fignk} is the nearly full suppression of the dispersion, $n(\nu)$, and absorption, $\kappa(\nu)$, for the backward propagation direction. In view of Eq.~(\ref{eqnk}) for this set of parameters the magnetoelectric susceptibility almost completely suppresses the electromagnon contribution.
To analyze this behavior in some more detail we modify Eq.~(\ref{eqnk}) taking into account the frequency dependencies of susceptibilities as given in the Appendix. We further assume in the first approximation that $\Delta\varepsilon|L_F(\omega)| \ll \varepsilon_{\infty}$, $\Delta \mu|L_F(\omega)| \ll 1$, $\Delta \mu \ll \Delta\varepsilon$, and carry out the calculations to the linear order of $L_F(\omega)$ only. Here $\Delta\varepsilon$ and $\Delta \mu$ are electric and magnetic contributions of the electromagnon, respectively; $L_F(\omega)$ is the Lorentzian function and $\varepsilon_{\infty}$ is the high-frequency dielectric permittivity. The final approximate expression may be written as:
\begin{equation}\label{eqnk2a}
  n^{\pm} \approx \sqrt{\varepsilon_{\infty}} + L_F(\omega) \frac{\sqrt{\varepsilon_{\infty}}}{2} \left(\sqrt{\frac{\Delta\varepsilon}{\varepsilon_{\infty}}} \pm \sqrt{\Delta \mu}\right)^2
\end{equation}
From Eq.~(\ref{eqnk2a}) we see that the effect of the magnetoelectric susceptibility on the electromagnon is equivalent to a weighted sum or difference of electric and magnetic contributions. Therefore, we expect complete suppression of the electromagnon contribution if the condition
\begin{equation}\label{eqnk3}
\Delta\varepsilon / \varepsilon_{\infty} \approx  \Delta \mu
\end{equation}
is fulfilled. As $\Delta\varepsilon$ is large for small magnetic fields ($\Delta\varepsilon \sim 30$) and is suppressed quadratically in high fields $\Delta\varepsilon \sim 1/B^2$~[\onlinecite{kuzmenko_prb_2014}], some optimum value of the electromagnon suppression is expected in moderate fields, as observed in Figs.~\ref{figtr},~\ref{fignk}. In such cases the directional anisotropy in \Sm~ is the strongest.
\begin{figure}[tbp]
\begin{center}
\includegraphics[width=0.85\linewidth, clip]{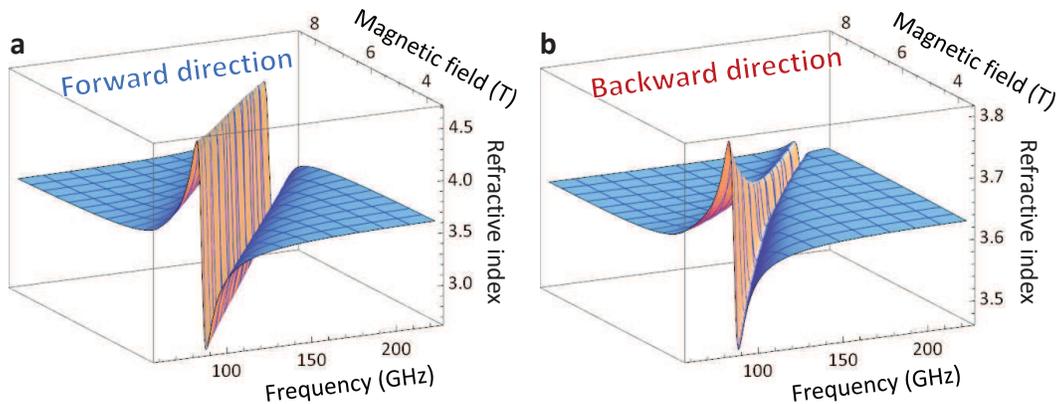}
\end{center}
\caption{\emph{Model calculations of directional birefringence}.
\textbf{a} - Refractive index for forward propagation direction for typical range of the parameters in \Sm. \textbf{a} - Same for backward propagation. Note the factor $8$ difference in scale between \textbf{a} and \textbf{b}. } \label{figth}
\end{figure}

In order to prove the estimates in Eq.~(\ref{eqnk3}) by rigorous calculations, we plot in Fig.~\ref{figth} the model curves for the refractive index in \Sm~ for forward and backward propagation directions, respectively. Typical absolute values of the characteristic parameters used in Fig.~\ref{figth} have been taken from the fits to the data above and from the results of previous experiments \cite{kuzmenko_prb_2014,kuzmenko_jetp_2011}. From the scale difference in Fig.~\ref{figth} we see that the amplitude of the Lorentzian is suppressed by at least a factor of $\Delta n^+ / \Delta n^- \sim 8$ in the full range of parameters. In agreement with estimates by Eq.~(\ref{eqnk3}) the suppression of the electromagnon for backward direction is the strongest for magnetic fields in the range $B \sim 5-8$~T, as observed in the experiment.

Compared especially with first experiments~\cite{krichevtsov_jetp_1988, krichevtsov_jpcm_1993, mukhin_ferr_1997} on non-reciprocal transmission in Cr$_2$O$_3$ the effect in samarium ferroborate is extremely strong and may approach absolute asymmetry in absorption. Such values are due to strong coupling of electric and magnetic order and is a characteristic feature of multiferroic materials with strong electromagnon~\cite{bordacs_nphys_2012, kezsmarki_nc_2014, kibayashi_nc_2014}. We note that in the data in Fig.~\ref{fignk}\textbf{a,b} no electromagnon mode is observed for backward polarization within experimental accuracy.

\section{Conclusions}

Directional birefringence and dichroism have been investigated in doped samarium ferroborate at millimeter-wave frequencies. We demonstrate strong directional anisotropy close to the resonance frequency of electromagnon excitation. The strength and position of the relevant electromagnon can be varied in external magnetic field. The non-reciprocity is demonstrated using two alternative approaches: i) by the inversion of the external magnetic field and ii) by explicit exchange of the source and detector. Nearly full suppression of the electromagnon in one propagation direction is observed. Approximate range of parameters to obtain strong directional anisotropy is investigated.

\subsection*{Acknowledgements}
This work was supported by Russian Foundation for Basic Researches (14-02-91000, 15-02-07647, 14-02-00307, 13-02-12442, 15-42-04186 ), Sci. school-924.2014.2 and by the Austrian Science Funds (I815-N16, I1648-N27, W1243).

\appendix
\section*{Appendix}

\subsection*{Symmetry analysis of the susceptibilities}
\label{app:A}

Pure and La-substituted SmFe$_3$(BO$_3$)$_4$ belong to R32 space group with easy-plain antiferromagnetic structure at low temperatures. At low temperatures and in external magnetic fields along the $y$-axis the following electric and magnetic configuration is stabilized~\cite{vasiliev_ltp_2006, kadomtseva_jetp_2012, kuzmenko_prb_2014, zvezdin_jetpl_2005, zvezdin_jetpl_2006, popov_jetp_2010}: $\mathbf{M} \|y$, $\mathbf{L} \|x$,  $\mathbf{P} \|x$. Here $\mathbf{M} \sim \mu_0 \mathbf{H}$ is the induced magnetization, $\mathbf{L}$ is the antiferromagnetic vector, and $\mathbf{P}$ is the static ferroelectric polarization. In this case the magnetic structure corresponds to $2'_x$ point group. From the symmetry arguments~\cite{birss_book} the following elements of the magnetoelectric susceptibility matrix are allowed: $\chi^{me,em}_{xy,yx,zx,xz}$. We note that in the dynamic case several additional terms (see below) may become nonzero.

Therefore, in order to obtain the relevant elements of the susceptibility matrices, a model of the static~\cite{zvezdin_jetpl_2005, zvezdin_jetpl_2006, popov_jetp_2010, mukhin_jetpl_2011} and dynamic~\cite{kuzmenko_jetpl_2011, kuzmenko_jetp_2011, kuzmenko_prb_2014}  properties  of \Sm ~has been utilized. The magnetoelectric energy
$\Phi_{me}(\vect{M}, \vect{L}, \vect{P})$ relevant for the present
analysis can be written
as~\cite{zvezdin_jetpl_2005,popov_prb_2013}:
\begin{equation}
\Phi_{me}(\vect{M}, \vect{L}, \vect{P}) = -c_1 \left[ P_x L_y L_z  -  P_y L_x L_z \right] -c_2 \left[ P_x \left(L_x^2 -
L_y^2 \right) - 2 P_y L_x L_y \right] + \mathellipsis \ .
\label{me_energy}
\end{equation}

The dynamic susceptibilities in \Sm~ are governed by two electromagnon modes \cite{kuzmenko_prb_2014,kuzmenko_jetpl_2011, kuzmenko_jetp_2011} which we denote as F-mode and AF-mode.
In the geometry $B \|b$ two modes are coupled and the susceptibility matrices $\hat{\chi}^m(\omega), \hat{\chi}^{me}(\omega), \hat{\chi}^{em}(\omega), \hat{\chi}^e(\omega)$  are obtained solving Landau-Lifshitz equations for dynamic magnetic variables $\Delta \mathbf{M}, \Delta \mathbf{L}$ similar to Ref.~[\onlinecite{kuzmenko_prb_2014}]:
\begin{equation}
\begin{array}{cc}
\hat{\chi}^m(\omega) \approx \left( \begin{array}{ccc}
\chi_{xx}^m & \chi_{xy}^m & \chi_{xz}^m \\
\chi_{yx}^m & \chi_{yy}^m & \chi_{yz}^m \\
\chi_{xz}^m & \chi_{zy}^m & \chi_{zz}^m \\
\end{array} \right)
& \hat{\chi}^{me}(\omega) \approx \left( \begin{array}{ccc}
\chi_{xx}^{me} & \chi_{xy}^{me} & 0 \\
\chi_{yx}^{me} & \chi_{yy}^{me} & 0 \\
\chi_{zx}^{me} & \chi_{zy}^{me} & 0 \\
\end{array} \right) \\[3em]
\hat{\chi}^{em}(\omega) \approx \left( \begin{array}{ccc}
\chi_{xx}^{em} & \chi_{xy}^{em} & \chi_{xz}^{em} \\
\chi_{yx}^{em} & \chi_{yy}^{em} & \chi_{yz}^{em} \\
0 & 0 & 0 \\
\end{array} \right)
& \hat{\chi}^e(\omega) \approx \left( \begin{array}{ccc}
\chi_{xx}^e & \chi_{xy}^e & 0 \\
\chi_{yx}^e & \chi_{yy}^e & 0 \\
0 & 0 & \chi_{zz}^e \\
\end{array} \right), \\
\end{array}
\label{susceptibilities1}
\end{equation}
where the components $\chi_{xy,yx,xz,zx}^m, \chi_{xx,yy,zy}^{me}, \chi_{xx,yy,yz}^{em}, \chi_{xy,yx}^{e}$ are of purely dynamic origin and they are proportional to $i\omega$. The absence of the components $\chi_{xz,yz,zz}^{me}, \chi_{zx,zy,zz}^{em}, \chi_{xz,yz,zx,zy}^{e}$ results from neglecting of a higher-order magnetoelectric term $P_z L_z L_x (L_x^2-3L_y^2)$.

The AF mode has been detected close to 320 GHz at low temperatures \cite{kuzmenko_jetpl_2011}. The resonance frequency of the F-mode is close to zero ($\sim 5$GHz) but increases roughly linearly in external magnetic fields \cite{kuzmenko_prb_2014}. In case which is relevant for the present experiment, the interaction between two electromagnon modes can be neglected. In this approximation and in the vicinity to the resonance frequency $\omega_F$ the susceptibility matrices may be simplified to the form given in Eq.~(\ref{susceptibilities}) of the main text.
%

% This approximation has been used throughout this work.

\subsection*{Data processing and transmission spectra}
\label{app:B}

We start from Maxwell equations which for plane waves ($\mathbf{E},\mathbf{H} \sim e^{i(\omega t - \vect{k}\cdot \vect{r})}$) can be written as:
\begin{eqnarray}
  i\vect{k}\times\vect{E} &=& i\omega \mathbf{B} = i \omega [ \mu_0(\hat{1}+\hat{\chi}^m) \vect{H} + \sqrt{\varepsilon_0 \mu_0}\hat{\chi}^{me} \vect{E}] \ , \label{eqmaxwell1}\\
 -i\vect{k}\times\vect{H} &=& i\omega \mathbf{D} = i \omega [ \sqrt{\varepsilon_0 \mu_0} \hat{\chi}^{em} \vect{H} + \varepsilon_0 (\hat{1}+\hat{\chi}^e) \vect{E}] \ .\label{eqmaxwell2}
\end{eqnarray}
Here $\hat{1}$ is the identity matrix.

In general, Eqs.~(\ref{eqmaxwell1},\ref{eqmaxwell2}) can be reduced to eigenvalue and eigenvector problems giving the wavevectors and the amplitudes of the propagating waves.

The electromagnetic field inside
the sample has the form of a plane wave $\exp(i (k_x x + k_y y + k_z
z - \omega t))$. The values of $k_x$ and $k_y$ are conserved at the
boundaries and they are determined by the geometry of the problem.
The value of $k_z$ depends on the properties of the sample and is
obtained solving the Maxwell equations within the sample following the Berreman
method~\cite{berreman_josa_1972}. Further details of the calculations may be found in Refs.~[\onlinecite{shuvaev_sst_2012,kuzmenko_prb_2014}]. In present experiments the light propagates along the $z$-direction, i.e. the $\mathbf{k}$ vector may be written as $\mathbf{k}=(0,0,k)$.

The solutions of Maxwell equations, Eqs.~(\ref{eqmaxwell1},\ref{eqmaxwell2}), are given by two linearly polarized modes. One solution with $\mathbf{h}\|y, \mathbf{e}\|x$ is trivial and reveals no dependence on the propagation direction:
\begin{equation}\label{eqnk2}
   k_1^{\pm} \lambda / 2 \pi=\pm \sqrt{\varepsilon_x \mu_y}
\end{equation}
Here $\varepsilon_x = 1+ \chi_{xx}^e$ is the dielectric permittivity along the $x$-axis and $\mu_y= 1+ \chi_{yy}^m$ is the magnetic permeability along the $y$-axis.

The propagation constant for a perpendicular polarization with $\mathbf{h}\|x, \mathbf{e}\|y$ does show non-reciprocal effects and it reads:
\begin{equation}\label{eqnk1}
  k_2^{\pm} \lambda / 2 \pi =  \tilde{\alpha}_{xy} \pm \sqrt{\tilde{\varepsilon}_y \tilde{\mu}_x}
\end{equation}
Here $\tilde{\alpha}_{xy} = \chi_{xy}^{me} -\chi_{zy}^{me} \chi_{xz}^{m} / \mu_{z}$, $\tilde{\varepsilon}_y = \varepsilon_y + (\chi_{zy}^{me})^2/\mu_{z}$, and $\tilde{\mu}_x = \mu_x + (\chi_{xz}^{m})^2/\mu_{z}$ are the renormalized material parameters;
$\varepsilon_y = 1+ \chi_{yy}^e$ is the dielectric permittivity along the $y$-axis,  $\mu_x= 1+ \chi_{xx}^m$ and $\mu_x= 1+ \chi_{xx}^m$ are the magnetic permeabilities along the $x$-axis and $z$-axis, respectively. In present approximation $\mu_x \approx \mu_z$. For a single-domain sample, the relevant electrodynamic parameters in Eq.~(\ref{eqnk1}) are obtained as \cite{kuzmenko_prb_2014}:
\begin{eqnarray*}
   \varepsilon_y &=& 1+ \chi_{yy}^e = \varepsilon_{\infty} + \Delta\varepsilon \cdot L_F(\omega) \\
   \mu_x &=& 1+ \chi_{xx}^m = 1 + \Delta\mu \cdot L_F(\omega)  \\
   \chi_{xz}^m &=& (i\omega/\omega_0) \Delta\mu \cdot L_F(\omega) \\
   \chi_{xy}^{me} &=& \rho \sqrt{\Delta\mu\Delta\varepsilon} \cdot L_F(\omega) \\
   \chi_{zy}^{me} &=& -i \frac{\omega}{\omega_F} \sqrt{\Delta\mu\Delta\varepsilon} \cdot L_F(\omega)
   \end{eqnarray*}
Here $\varepsilon_{\infty}$ is the high-frequency contribution to the permittivity, $\omega=2\pi \nu$ is the angular frequency. $\omega_0$, $\Delta\mu$, and $\Delta\varepsilon$ are the resonance frequency, magnetic, and dielectric contribution of the low-frequency electromagnon, respectively. The factor $\rho(H) = H / \sqrt{H^2 + 2 H_A' H_E}$ reflects the
changes of the magnetic structure with increasing magnetic field and it
becomes unity in the fields exceeding 5-10~kOe. Here $H_A'$ and $H_E$ are anisotropy and exchange fields, respectively.

We utilize the Lorentzian form for the electromagnon resonance with
\begin{equation}
   L_{F}(\omega) = \omega_{F}^2 / \left(\omega_{F}^2 - \omega^2 + i \omega g \right) \ , \nonumber
\end{equation}
with $g$ being the resonance width.

The normalized impedance of the medium for the solution given by Eq.~(\ref{eqnk1}),  $z \sqrt{\mu_0 /\varepsilon_0}=E_y/H_x$, does not depend upon the propagation direction and may me written as:
\begin{equation}\label{eqz}
 z= \sqrt{{\tilde{\mu}_x}/{\tilde{\varepsilon}_y}} \ .
\end{equation}

The transmission amplitude in case of a plane parallel sample can be calculated explicitly as
\begin{equation}\label{eqtr}
 t^{\pm}= \frac{(1-r^2)e^{ik^{\pm} d}}{1-r^2 e^{i(k^{+}+k^{-}) d} } \ .
\end{equation}
Here $r=(z-1)/(z+1)$ is the reflectivity on the sample surface.

We utilized Eq.~(\ref{eqtr}) to obtain the fit curves in Figs.~\ref{figtr},\ref{fignk}. We note that in Eq.~(\ref{eqtr}) only the term $e^{ik^{\pm} d}$ is asymmetric with respect to the propagation direction. This means that relative birefringence and dichroism may be directly obtained from the ratio $t^+ / t^-$ without knowing the details of the experiment.
In order to rigorously invert the transmission data in Fig.~\ref{figtr}, the values of $r$ are missing in Eq.~(\ref{eqtr}). However, as we can get from the fitting procedure, $r(\omega)$ varies only by $\pm 3\%$ in the full range and by $\pm 0.2\%$ far from the resonance. Therefore, and without loosing the accuracy, $r(\omega)$ can be chosen as a constant. In order to improve the approximation, $r(\omega)$ has been taken from the fits to the transmission data in Fig.~\ref{figtr}. With known $r(\omega)$ the inversion of Eq.~(\ref{eqtr}) and the calculation of $k^+$ and $k^-$ from $t^+$ and $t^-$ is straightforward. The results of this procedure are shown as symbols in Fig.~\ref{fignk}.

\bibliography{literature}

\end{document}